\documentclass[a4paper,11pt]{article}

\title{\vspace{-2.5cm} \bf On the fate of Birkhoff's theorem in Shape Dynamics}
\author{Flavio Mercati\footnote{fmercati@perimeterinstitute.ca.},
\vspace{12pt} \\
\it \small Perimeter Institute for Theoretical Physics,\\
\it \small 31 Caroline Street North,
Waterloo, ON, N2L 2Y5 Canada,}
\date{\today}

\usepackage{amssymb,bbold}
\usepackage{amsmath}
\usepackage[usenames,dvipsnames]{color}
\usepackage{tensor}
\usepackage{bbm}
\usepackage{graphicx}
\usepackage{cancel}
\usepackage[left=1in,right=1in,top=1in,bottom=1in]{geometry}                  
\usepackage{setspace}
\usepackage{hyperref}
\usepackage{color}
\usepackage{wrapfig}
\usepackage{lipsum}
\usepackage{mathrsfs}

\hypersetup{
    colorlinks=true,         
    linkcolor=blue,          
    citecolor=red,        
    urlcolor=Violet             
}

\newcommand{\sfrac}{\textstyle \frac}
\newcommand{\st}[1]{\text{\tiny \rm #1}}

\renewcommand{\d}{{\rm d}}

\newcommand{\wid}{0.48\textwidth}
\newcommand{\widfull}{0.6\textwidth}

\begin{document}

\maketitle

\begin{abstract}
Spherically symmetric, asymptotically flat solutions of Shape Dynamics were previously studied assuming standard falloff conditions for the metric and the momenta. These ensure that the spacetime is asymptotically Minkowski, and that the falloff conditions are Poincar\'e-invariant. These requirements however are not legitimate in Shape Dynamics, which does not make assumptions on the structure or regularity of spacetime. Analyzing the same problem in full generality, I find that the system is underdetermined, as there is one function of time that is not fixed by any condition and appears to have physical relevance. This quantity can be fixed only by studying more realistic models coupled with matter, and it turns out to be related to the dilatational momentum of the matter surrounding the region under study.
\end{abstract}

\section{Introduction}

In this paper, I study in complete generality spherically symmetric, asymptotically flat vacuum solutions of Shape Dynamics (SD). This is a theory of gravity that replaces spacetime for a history of conformal 3-geometries~\cite{FlaviosSDtutorial}. The theory comes equipped with a procedure to deduce a 4D spacetime line element (foliated by a Constant-Mean-extrinsic-Curvature (CMC) slicing) from each of its classical solutions, which has been developed in the 70's by J.~York and others~\cite{York1971,ChoquetBruhat73,Niall_73}. Such a 4D metric has the physical meaning of the spacetime that is `experienced' by weak matter fluctuations propagating on the SD solution. In principle, such a 4D metric does not have to satisfy Einstein's equations, and does not necessarily have to be regular (it is the 3D conformal geometry that has to be well-behaved). This opens up the possibility of interpreting spacetime singularities as `artifacts' of the spacetime description, which are absent in the description as evolving conformal geometry. 
In principle, the dynamics of SD could be generated by any conformally (and diffeomorphism) -invariant Hamiltonian, and in the spirit of effective field theory one would like to allow for the most generic dynamics, and study the flow of the free parameters under renormalization. This, however, is a long-term program which requires a significant amount of preparatory work, in particular a good understanding of classical SD. The approach that has been followed so far is to start from a certain `subtraction point' of the RG flow, which corresponds to the astrophysical/cosmological regime where GR can be currently tested. There, we can fix our Hamiltonian to be the one that gives a dynamics that is equivalent to that of GR (York's method~\cite{York1971,ChoquetBruhat73,Niall_73} allows us to calculate such a Hamiltonian, which is nonlocal, as the solution of an elliptic quasilinear differential equation). The resulting theory is locally identical to the Arnowitt--Deser--Misner formulation of General Relativity in a CMC foliation. Such a local equivalence, however, does not imply a complete equivalence between the two theories: it turns out that the fundamental requirement of SD, namely, that the conformal geometry be regular throughout each solution, translates into solutions which cannot be globally interpreted as spacetimes.

This is clearly seen in the case studied in~\cite{Birkhoff_SD}, which provides the starting point for the present paper: in the spherically symmetric, asymptotically flat case SD admits solutions which locally look like Schwarzschild's spacetime, but globally can only be interpreted as peculiar choices of patches of Kruskal's extension. Specifically, in the solution of~\cite{Birkhoff_SD} the constraint equations admit solutions with a spatial metric that becomes Lorentzian within a certain region. These are perfectly acceptable in GR (in Schwarzschild coordinates the radial coordinate becomes timelike inside of the horizon), but a change of signature of the metric represents a discontinuity of the conformal geometry, and therefore is unacceptable in SD. The problem is that one is using an unattainable choice of  spatial coordinates. In fact, by closer inspection of the equations, one sees that there are limitations to the diffeomorphism gauge choice, which, if respected, ensure the regularity of the conformal geometry: the areal radius of the metric ($\sqrt{g_{\theta\theta}}$) has to respect a bound, which represents a `throat': such a metric cannot accommodate spheres of area smaller than a certain value. Using a gauge choice that respects this bound, one ends up with a metric that is continuous, has a regular conformal part, and two asymptotically flat ends. Topologically such a metric is that of a wormhole, and has no singularity.

The results of~\cite{Birkhoff_SD}, however, depend strongly on the assumption of asymptotic flatness. Such an assumption is standard in the literature on GR, but SD's foundations call for a critical reassessment of its validity. In fact Shape Dynamics is entirely well-defined only in the spatially compact case, where, as Einstein put it, ``the chain of cause and effect is closed''~\cite{EinsteinChainOfCauses}. This reflects in the fact that non-compact-space solutions depend on arbitrary boundary conditions, while if space is compact all boundary conditions are completely determined by the topology of space, and the system is genuinely self-contained. From the perspective of Shape Dynamics, a noncompact spatial slice is a conformal geometry with a piercing, corresponding to the point at infinity, and we are free to choose what the fields do at this piercing. To have a good understanding of asymptotic flatness, we should study some simple compact solutions and try to understand in which regimes there are empty regions of such solutions that can be well approximated by asymptotically flat spaces. In other words, physically-realistic compact solutions should provide, dynamically, the boundary conditions that we need in the asymptotically-flat case. 

This problem will be addressed in future works. Before that, we need to understand what kind of asymptotic boundary conditions SD allows in the spherically symmetric case. In GR, this question is answered by Birkhoff's theorem: there is only one vacuum, spherically symmetric solution of Einstein's equations. Any spherically symmetric vacuum solution one might find is bound to be isometric to Schwarzschild's line element. But SD is a theory with a preferred notion of simultaneity, corresponding, where the equivalence with GR is valid, to a CMC hypersurface.\footnote{In the asymptotically flat case it is appropriate to take a \emph{maximal} hypersurface (which is when the extrinsic curvature is zero). Such a foliation can be seen as a small interval of CMC leaves. The analysis of~\cite{Birkhoff_SD} was made in such a foliation.}
It is not obvious whether something like Birkhoff's theorem holds in SD: there might be more than one spherically-symmetric vacuum solution of the theory, and they would correspond to different maximal slicings of Schwarzschild's spacetime.

The result of~\cite{Birkhoff_SD} relies on some asymptotic falloff conditions for the 3-metric $g_{ij}$ and its conjugate  momenta $p^{ij}$ which are well-justified only in GR. In particular the momenta   are assumed to falloff at infinity like $1/r^2$. This is a standard assumption in the Hamiltonian approach to GR, and follows from requiring Poincar\'e invariance of the asymptotic falloff conditions of the metric and momenta~\cite{beig1987poincare}. This assumption is however too restrictive for SD: our \emph{spatial} slices have to be asymptotically flat, but we do not have to require that the \emph{spacetime} geometry asymptotes to Minkowski. In this paper, I make no assumption and study the most general case. What I find is that there is no equivalent of Birkhoff's theorem in SD: the solution is not unique, as there is a single function of time $A=A(t)$ that no equation fixes. This function can only be fixed  in a dynamically closed - and therefore spatially compact - universe: it is related (through the diffeomorphism constraint) to the dilatational momentum of matter surrounding the empty, spherically-symmetric region under consideration. The system is underdetermined, as any choice of $A(t)$ gives an equally-valid solution of the equations. Of this family of solutions, only the case $A =0$ gives Lorentz-invariant falloff conditions whose momentum falls off like $1/r^2$. This was the choice made in~\cite{Birkhoff_SD}. All other choices introduce a preferred frame of reference, which can be identified by the state of motion of the matter at infinity.

\section{Solution of the constraints}

SD is a gauge theory of conformal and spatial diffeomorphism symmetry. This means that, although the only physical degrees of freedom in the theory are conformally and diffeomeorphism-invariant, we still use a redundant description which depends on a particular choice of coordinates and of conformal gauge. There is a practical reason for this: the gauge-invariant degrees of freedom are nonlocal, while a redundant description can be local.
In particular, by choosing the conformal gauge in which the theory is locally equivalent to GR (ADM in CMC or maximal slicing) we can avoid the problem of dealing with a nonlocal Hamiltonian. In this paper, in particular, I will work in this gauge, and will be studying solutions of Arnowitt--Deser--Misner gravity in a maximal foliation. When such solutions exist, they are both solutions of GR and SD. However there are situations in which such solutions do not correspond to a well-defined solution of Einstein's equations, in particular at the Big-Bang singularity~\cite{ThroughTheBigBang}. However, by looking at the conformally-invariant degrees of freedom, one can check whether, as solutions of SD, 
they still make sense and can be continued past such breakdown point. My strategy is clear: I want to work with ADM gravity in CMC/maximal gauge as long as it is possible, and then focus on the shape degrees of freedom when my solutions evolve into something that cannot be described in GR.

Let us begin by writing the constraint equations of ADM gravity in maximal slicing:
\begin{equation}
\begin{aligned}
&\mathcal H = {\sfrac{1}{\sqrt g}} \left( p^{ij} p_{ij} - {\sfrac 1 2} p^2 \right) - \sqrt g \, R  \approx 0 \,,&
&\mathcal H_i  = -2 \, \nabla_j p^j{}_i  \approx 0 \,,&
&\mathcal C  = g_{ij} p^{ij}  \approx 0 \,.&\label{Constraints}
\end{aligned}
\end{equation}
$\mathcal H$ is the Hamiltonian constraint, $\mathcal H_i $ is the diffeomorphism constraint and $\mathcal C$ is the conformal constraint, that imposes that the momentum is traceless. This, in the spacetime description, implies that the extrinsic curvature of my foliation (which is related to the momentum by $K_{ij} = \frac{1}{\sqrt g} \left( \frac 1 2 g_{k\ell} p^{k\ell} g_{ij} -  p_{ij} \right)$) has zero trace. In SD, however, the interpretation of $\mathcal C$ is different: it is the generator of conformal transformations (as can be verified by taking its Poisson brackets with the metric and the momentum), while the Hamiltonian constraint $\mathcal H$ plays the role of a gauge-fixing of $\mathcal C$ (in fact the two are second-class). This gauge-fixing selects the conformal gauge in which SD and GR are equivalent.

Assuming a spherically symmetric ansatz
\begin{equation} \label{SphericalSymmetry}
\d s^2 =  \mu^2 \d r^2 +  \sigma \left( \d \theta^2 + \sin^2 \theta \d \phi^2 \right) \,,
\qquad
p^{ij} =  \text{diag} \, \left\{  \frac{f}{\mu} , {\sfrac 1 2}  s   , {\sfrac 1 2}  s  \, \sin^{-2} \theta  \right\} \, \sin \theta \,,
\end{equation}
the constraints~(\ref{Constraints}) take the form (the sign $'$ stands for derivative with respect to $r$):
\begin{eqnarray}
&&\frac 1 {2   \sigma   \mu^2} \left[ 2 f   \sigma   \mu^2   s - f^2   \mu^3 + \mu (\sigma')^2 + 4 \sigma   \mu^3 + 4 \sigma \mu' \sigma'  - 4 \sigma \mu \sigma'' \right] = 0  \,, \label{HamConstEq}
\\
&& \mu f' - {\sfrac 1 2} s \sigma' = 0\,,  \label{DiffeoConstEq}
\\
&&\mu f + s \sigma = 0 \,,  \label{ConfConstEq}
\end{eqnarray}
we take the last constraint,~(\ref{ConfConstEq}), to fix $s = - \mu f /\sigma$. The diffeo constraint(\ref{DiffeoConstEq}) then reads
\begin{equation}\label{SolDiffeoConstraint}
(f \sqrt \sigma)' = 0 \,, \qquad \Rightarrow \qquad 
f =  A /{\sqrt \sigma} \,,
\end{equation}
where $A$ is an integration constant. 
Now, the Hamiltonian constraint~(\ref{HamConstEq}) can be written as
\begin{equation} \label{SecondFormOhHamConst}
2 \left[ \left( \frac{\sigma'} \mu \right)' - \mu - {\sfrac 1 4} \frac{(\sigma')^2}{\mu \sigma} + {\sfrac 3 4} \sfrac{\mu f^2}{\sigma} \right] =0 \,,\end{equation}
and, calling $\varphi =  \frac{\sigma'}{\sigma^{\frac 1 4} \mu} $,~(\ref{SecondFormOhHamConst}) can be rewritten
\begin{equation}
\frac{\sigma^{\frac 1 4}}{2 \varphi} \left( \varphi^2 - 4 \sqrt{\sigma} - \frac{f^2}{\sqrt{\sigma}} \right)' 
 =  - f \mu \left(\frac{f'}{\sigma'} + {\sfrac 1 2} \frac f \sigma   \right) \,,
\end{equation}
which, inserting the solution~(\ref{SolDiffeoConstraint}) and multiplying by $2 \varphi \sigma^{-1/4}$, reads
$ \left( \varphi^2 - 4 \sqrt{\sigma} - \frac{f^2}{\sqrt{\sigma}} \right)' = 0$. This gives us a first integral of Equation (\ref{SecondFormOhHamConst}), which we call $m$:
\begin{equation}\label{DefinitionOfm}
\varphi^2 - 4 \sqrt{\sigma} -  f^2/\sqrt{\sigma}  = - 8 \, m \,.
\end{equation}
Using the above and the solution to the diffeo constraint (\ref{SolDiffeoConstraint}), Equation  (\ref{SecondFormOhHamConst}) turns into
\begin{equation} \label{ThirdFormOhHamConst}
\sigma (\sigma')^2 = \left( 4 \sigma^2 - 8 m \sigma^{\frac 3 2} + A^2 \right) \mu^2 \,
\,.
\end{equation}
The above equation is compatible with asymptotically flat boundary conditions for the metric
$\sigma = r^2 +\mathcal{O}(r)$, $\mu = 1 +\mathcal{O}(r^{-1})$. 
Moreover, the integration constant $m$ coincides with the ADM mass of the system:
using Gourgoulhon's definition~\cite{Gourg} (p. 163) and Eq.~(\ref{ThirdFormOhHamConst}):
\begin{equation}
M_\st{ADM}  = 
\frac 1 2  \left( r \mu^2 - \sigma' + \sigma/r \right)  = \frac 1 2 \lim_{r\to \infty} \left( \frac{r \sigma (\sigma')^2}{4 \sigma^2 - 8 m \sigma^{\frac 3 2} + A^2} - \sigma' + \sigma/r \right) = m \,.
\end{equation}
In Sec.~\ref{EqOfMotionSubSec}, where I calculate the equations of motion, I first show that $m$ is a conserved quantity, and then that its definition (\ref{DefinitionOfm}), considered as a functional of the metric and momentum, coincides with the Misner--Sharp mass (as defined, \emph{e.g.} in~\cite{GiuliniMcVittie}).

\subsection{The different phases}
\begin{figure}[h]
\begin{center}
\includegraphics[width=\widfull]{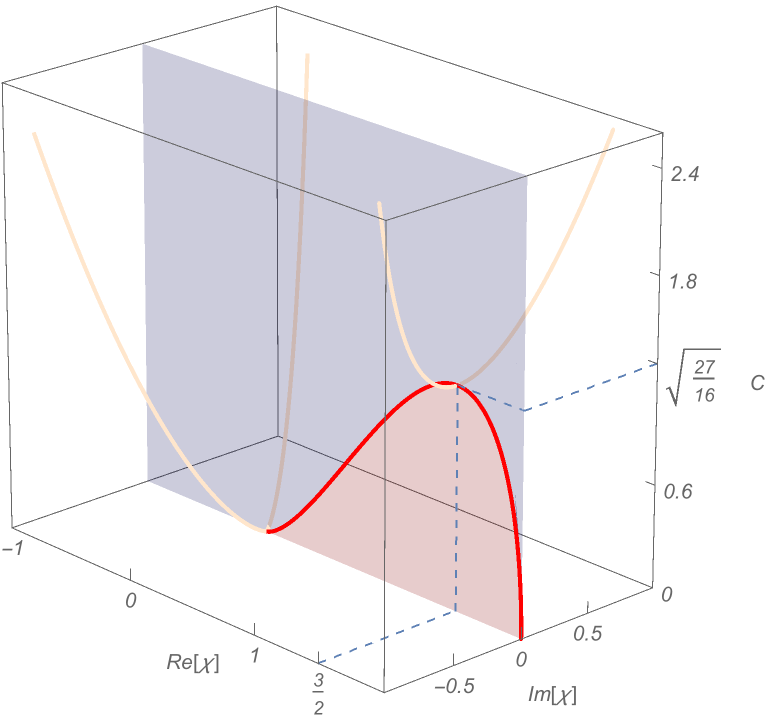}
\end{center}
\caption{\small Plot of the four roots of the polynomial $\chi^4 - 2 \chi^3 + C^2$ (positive-$m$ case). The horizontal axes represent the real and imaginary components of the root. The vertical axis represents the value of $C$. The blue region in the $\chi \in \mathbb{R}$ plane is the region in which $\chi^4 - 2 \chi^3 + C^2>0$, and the red region is the one in which the polynomial is negative.}
\label{Phases}
\end{figure}

Eq.~(\ref{ThirdFormOhHamConst}) has to be treated differently depending on whether the right-hand side is always positive. Whatever the values of $m$ and $A^2$, there exists a $\bar \sigma >0$ such that the quantity $ \left( 4 \sigma^2 - 8 m \sigma^{\frac 3 2} + A^2 \right)$ will be positive $\forall \sigma > \bar \sigma $. Depending on the values of $m$ and $A$, however, this quantity can have negative values over a range of $\sigma$. Switching to dimensionless variables $\chi^2 = \sigma/m^2$ and $C = A/(2 m^2)$, the quantity turns into the following polynomial:
\begin{equation}
\mathscr P(\chi) = ( \chi^4 - \text{sign}(m) \, 2 \,\chi^3 + C^2 ) \,.
\end{equation}
$\mathscr P $ has no positive roots if $m$ is negative.\footnote{If $m<0$, $\mathscr P (\chi)$ has two complex and two real negative root for $C^2 \in [0,27/16)$, and four complex roots for $C^2 > 27/16$.} 
In the positive-ADM-mass-case, $\text{sign}(m) = +1$, then there will be two positive and two complex conjugate roots, for any $C^2 \in [0,27/16)$. For $C^2 = 27/16$ the two real roots collapse to a double root, $\chi=3/2$. For $C^2 > 27/16$ all roots are complex. In Figure~\ref{Phases} below I plotted the four roots of the polynomial as functions of $C$.

Notice that the ADM mass has been proved to be positive if matter satisfies the dominant energy condition by Schoen and Yau in 1979~\cite{schoen1979proof}. Such proof, and the alternative proofs like that of Witten~\cite{witten1981new} of course assume the standard asymptotically flat falloff conditions $p_{ij} \sim 1/r^2$.  Therefore in our context they should be revised, because the are not valid when $A\neq 0$. For now, I will postpone such a discussion to a dedicated work, and I will remain agnostic regarding the possible signs of $m$.

\subsubsection{The singular phase}\label{SingularPhaseSec}

If $C^2 > 27/16$ with positive ADM mass, or if the ADM mass is negative or zero, then the polynomial showed above is always positive (for positive $\chi$). In this case we can use the areal radius $y^2 = \sigma$ as a radial coordinate:
\begin{equation}\label{FourthFormOhHamConst}
\d s^2 =\frac{y^4}{ y^4 - 2 \, m \, y^3 + A^2/4} \d y^2 + y^2 \left( \d \theta^2 + \sin^2 \theta \d \phi^2 \right) \,.
\end{equation}
\def\figsubcap#1{\par\noindent\centering\footnotesize(#1)}
\begin{figure}[h]%
\begin{center}
 \parbox{\wid}{\includegraphics[width=\wid]{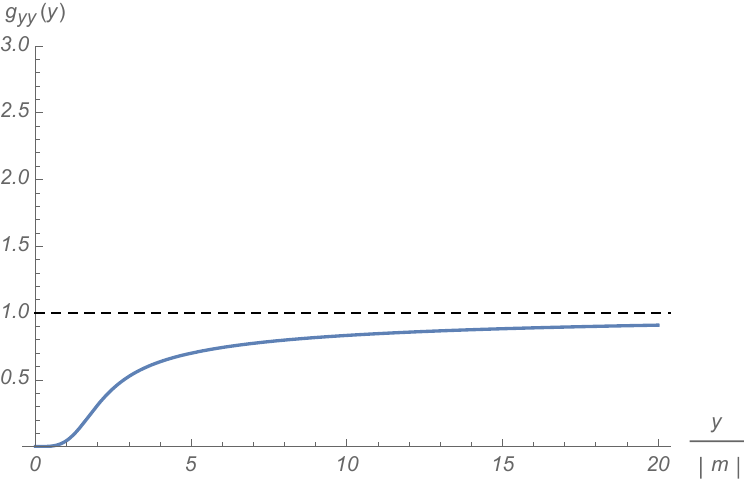}\figsubcap{a}}
 \hspace*{4pt}
 \parbox{\wid}{\includegraphics[width=\wid]{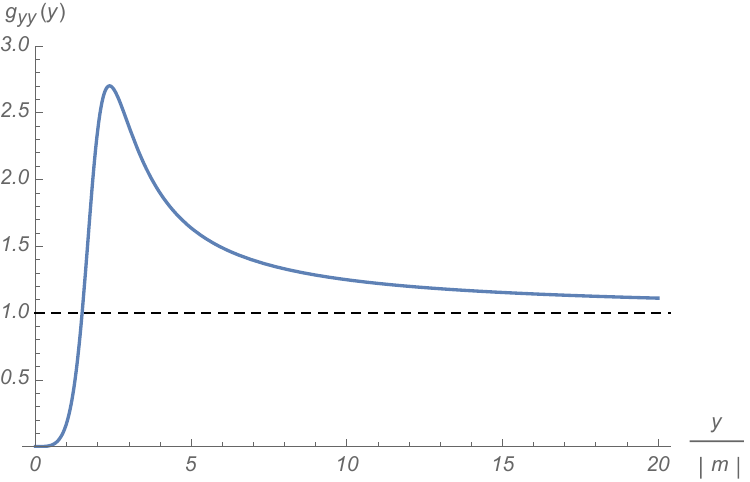}\figsubcap{b}}
 \caption{\small $y$-$y$ component of the metric in the singular phase.
 (a) If the ADM mass is negative. (b) If $m>0$ but  $C^2 > 27/16$. The maximum is always at $y=3/2$, and its value diverges as $C^2 \to (27/16)^+$.}
\label{Phase1}
\end{center}
\end{figure}
The metric~(\ref{FourthFormOhHamConst}) has some interesting features. First of all, the first derivative of the $y$-$y$ component at $y=0$ depends on the value of $A$:
\begin{equation}
\partial_y g_{yy} = \frac{16 y^3 \left(A^2-2 m y^3\right)}{\left(A^2+4 y^3 (y - 2m)\right)^2} \,,
\qquad  \lim_{y\to 0} \partial_y g_{rr}  = \left\{ \begin{array}{lll}
0  & \text{\it if} & A\neq 0 \\
- \frac{1}{2m} & \text{\it if} & A= 0
\end{array}\right. \,.
\end{equation}
So, if $A=0$, one has a conical singularity at the origin. However the 3-geometry is always singular at the origin. In fact the three independent curvature invariant \emph{densities} take the form:
\begin{equation}
\begin{aligned}
&\sqrt g \, R = \frac{3 A^2}{y^6 \sqrt{A^2+4 y^3 (y-2 m)}}\,,  ~~ \sqrt g \, R^{ij} R_{ij} = \frac{3 \left(3 A^4-8 A^2 m y^3+16 m^2 y^6\right)}{4 y^8 \sqrt{A^2+4 y^3 (y-2 m)}} \,,&\\
&\sqrt g \, R^{ij} R_{jk} R^k{}_i =  \frac{3 \left(11 A^6-60 A^4 m y^3+144 A^2 m^2-64 m^3 y^9\right)}{16 y^{14} \sqrt{A^2+4 y^3 (y-2 m)}} \,, &
\end{aligned}
\end{equation}
and some of them diverge near the origin, if either $A$ or $m$ are nonzero.
so we have a curvature singularity there. The conformal geometry, however, is regular everywhere: the Cotton--York tensor density is finite: $C^i{}_j = \epsilon^{ik\ell}  \left( \nabla_k R_{\ell j} - \sfrac 1 4 \nabla_k R g_{\ell j} \right) = 0$.

It is also important to remark that, if one constructs a 4-metric ${^\st{(4)}g}_{\mu\nu}$ by solving the lapse-fixing equation and the equations of motion for the metric (as is done in Sec.~\ref{EqOfMotionSec}), then  ${^\st{(4)}g}_{\mu\nu}$ solves everywhere Einstein's vacuum equations  (except, possibly, at the origin). All the 4D Riemannian invariants are those of a Schwarzschild metric with mass $m$, so, independently of the value of $A$, our metric in ADM gauge always covers some patches of Schwarzschild's spacetime.

\subsubsection{The wormhole phase}\label{WormholePhaseSec}

In the regime where the polynomial has real positive roots (when $C^2 \in [0,27/16)$ and $m>0$), the coordinate choice $\sigma = y^2$ is not good, as the right-hand side of Eq.~(\ref{ThirdFormOhHamConst}) becomes negative in some coordinate interval, which means that $\mu$ should become imaginary and the 3-metric Lorentzian.
In GR this is acceptable, but it is not in SD, where the conformal 3-geometry must be regular, and a change of signature is a discontinuity in the conformal 3-geometry. So we are forced to assume that $\sigma' = 0$ at that point, which is then an extremum. Since the metric has been assumed to be asymptotically flat at infinity, the extremum must be a minimum. This is a \emph{throat:} in this geometry one cannot fit any sphere with area smaller than $4\pi$ times the minimum of $\sigma$.

At this point, to make further progress we need to choose a diffeo gauge (a coordinate choice). The gauge choice is not completely arbitrary: it must be \emph{attainable,} in the sense that it must respect the conformal flatness and regularity of the conformal geometry. One choice is the isotropic gauge, $\mu = \sigma / r^2$. This choice translates the attainability condition into a requirement of regularity and positivity of the function $\sigma$. However, in this gauge, Eq.~(\ref{ThirdFormOhHamConst}) turns into an integral equation involving an elliptic integral for which there is no closed-form solution. It is convenient to use a different type of gauge, one in which the form of $\sigma = \sigma(r)$ is explicitly fixed. This form must respect the attainability conditions and the asymptotic flatness requirement. One of the simplest possible choice is (for positive $m$):
\begin{equation}\label{DiffeogaugeSigma}
\sigma = \left( 1 + \frac{m \chi_2}{4 r} \right)^4 r^2 \,,
\end{equation}
where $\chi_2 = \chi_2(C)$ is the largest positive root of the polynomial $\left( \chi^4 - 2 \chi^3 + C^2 \right)$. Such a function has a unique minimum at $r_\st{min} = m \chi_2/4$, corresponding to $\sigma (r_\st{min}) = m^2 \chi_2^2$. The corresponding expression for $\mu$ is deduced straightforwardly from Eq.~(\ref{ThirdFormOhHamConst}),
\begin{equation}\label{DiffeogaugeSigma2}
\mu^2 = \frac{r^4 \left(\frac{m \chi _2}{4 r}+1\right){}^8 \left(1 -\frac{m^2 \chi _2^2}{16 r^2}\right){}^2}{C^2+\frac{r^4}{m^4} \left(\frac{m \chi _2}{4 r}+1\right){}^8-\frac{2 r^3}{m^3} \left(\frac{m \chi _2}{4 r}+1\right){}^6 } \,.
\end{equation}
It is easy to verify that this form of $\mu$ is regular for any finite nonzero $r$. For large $r$, $\mu \to 1 + \mathcal{O}(1/r)$, while for small $r$ $\mu^2 \sim m^4 \chi_2^4/(256 r^4) + \mathcal{O}(1/r^3)$. Since in the same limit $\sigma \sim m^4 \chi_2^4/(256 r^2)$, we see that the metric tends to the image of an asymptotically flat metric under the inversion map $r \to \frac{m^2\chi_2^2}{16 r}$.
\begin{figure}[h]%
\begin{center}
 \parbox{\wid}{\includegraphics[width=\wid]{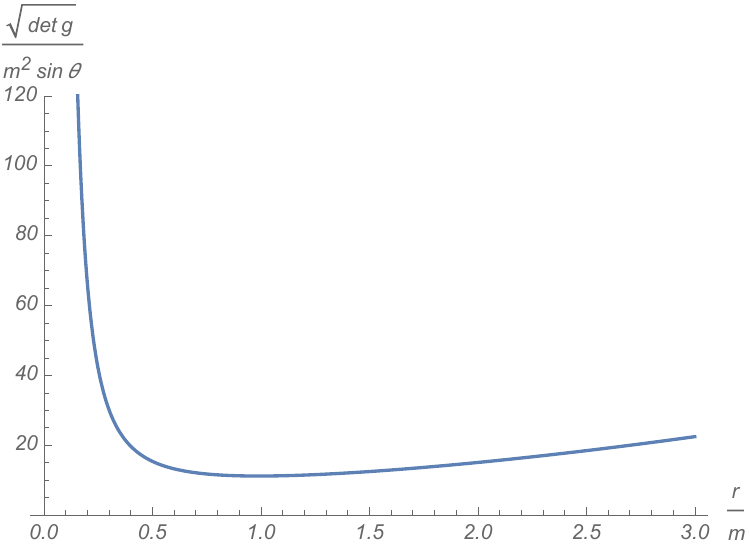}\figsubcap{a}}
 \hspace*{4pt}
 \parbox{\wid}{\includegraphics[width=\wid]{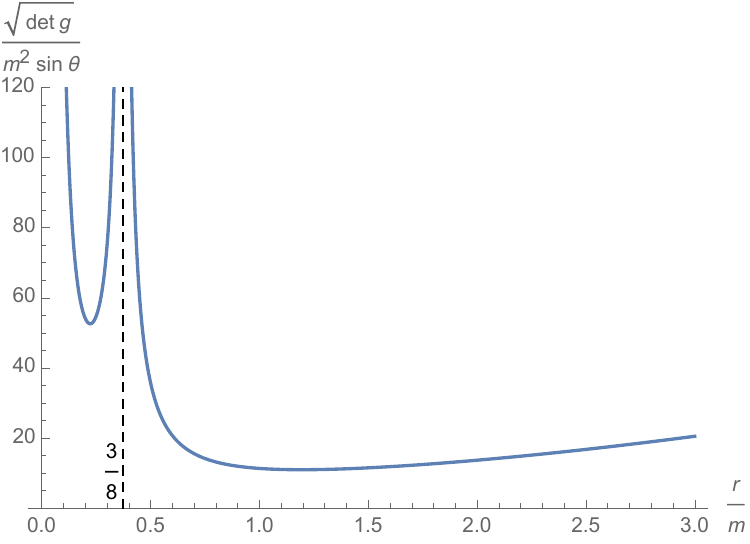}\figsubcap{b}}
 \caption{\small  Plot of the volume element $\sqrt{g} = \sqrt{\mu^2 \sigma^2}$ as a function of $r$ for all possible values of $C$. 
 (a) For $C^2 \in [0,27/16)$ it is a function that is positive everywhere, it has a minimum at a finite $r$ and goes to $r^2$ for large $r$ and to the corresponding inverted volume element $\sim r^{-4}$ for small $r$.  (b) When $C^2$ approaches $27/16$ from below $\sqrt{g} $ develops a singularity at $r = \frac 3 8 m$. At that value of $r$ the component $\sigma = 9/4$, and therefore $\sqrt{\sigma} = 3/2$, which corresponds to the value of $y$ where the metric  I studied above (in the case $C^2>27/16$, $m>0$) had a singularity in the limit $C^2 \to (27/16)^+$. So there is continuity between this case and the case studied above, when  $C^2$ approaches the critical value $27/16$.}
\label{Phase2}
\end{center}
\end{figure}

\subsubsection{Discussion of the two phases}

If we concentrate on the behaviour far away from the origin, in principle, both phases are physically acceptable, as they give rise to a regular conformal geometry in a neighborhood of the point at infinity (meaning everywhere except a compact ball centred around the origin). The only case that is likely to be excluded on physical grounds is that of negative ADM-mass, for which however we still do not have the SD equivalent of positive-energy theorems.

The behaviour close to origin is what distinguishes the two phases. In the `wormhole' phase the geometry has a throat, which means that  it cannot support concentric spherical surfaces of area smaller than a certain value. For algebraic reasons, the areal radius of the metric has to be monotonic except at the one point where it has its minimum, so it has to diverge both at $r \to \infty$ and at $r \to 0$. Then the region between the throat and $r =0$ is decompactified, and 
 the only solution of the problem in this case has two  asymptotically flat ends. This is all we  need to know regarding the metric. We will look at the behaviour of matter propagating on this background once we will have studied the equations of motion.

Regarding the singular phase, it is, indeed, singular: the areal radius is now monotonic all the way through from infinity to zero, and where it reaches zero we have a curvature singularity of the metric. The conformal geometry is everywhere regular, except at the origin which houses a puncture. Such a solution is physically acceptable as long as the singularity at the origin is shielded by some matter, which will change the effective values of the effective integration constants $m$ and $A$ inside its domain, so that they attain the only values that are compatible with a smooth, compact origin: zero. So the presence of the singularity cannot be used to rule out the corresponding values of $m$ and $A$, as long as there is some matter around the origin. In the absence of matter, these values of $m$ and $A$ can be definitely excluded.

\section{Solution of the equations of motion}\label{EqOfMotionSec}

\def\Xint#1{\mathchoice
   {\XXint\displaystyle\textstyle{#1}}%
   {\XXint\textstyle\scriptstyle{#1}}%
   {\XXint\scriptstyle\scriptscriptstyle{#1}}%
   {\XXint\scriptscriptstyle\scriptscriptstyle{#1}}%
   \!\int}
\def\XXint#1#2#3{{\setbox0=\hbox{$#1{#2#3}{\int}$}
     \vcenter{\hbox{$#2#3$}}\kern-.5\wd0}}
\def\ddashint{\Xint=}
\def\dashint{\Xint-}

To keep following my strategy of studying solutions of SD in the gauge in which there is equivalence with GR in a maximal slicing, I now need to study the equations of motion of ADM gravity that are generated by the Hamiltonian constraint smeared with the lapse that propagates the maximal-slicing condition. Such a lapse function can be obtained by looking at the kernel of the Poisson bracket between the constraints $\mathcal H$ and $\mathcal C$, which is a second-order differential operator. Finding this kernel requires to solve a linear elliptic differential equation:
\begin{equation}\label{LFE}
{\sfrac 1 2} \sqrt g N R - 2 \sqrt g  \Delta N + {\sfrac 3 2} \frac {p^{ij} p_{ij}}{\sqrt g} N = 0 \,,
\end{equation}
which is usually called \emph{lapse fixing equation.}
Once we have the kernel of $ \{ \mathcal H , \mathcal C \}$ we can use it in the ADM equations of motion for the metric:
\begin{equation}\label{gdotEq}
\dot g_{ij} = \frac{2 N}{\sqrt g} \left( p_{ij}- {\sfrac 1 2} g_{ij} p \right) + \nabla_i \xi_j + \nabla_j \xi_i \,,
\end{equation}
and for the momenta:
\begin{equation}\label{pdotEq}
\begin{aligned}
\dot p^{ij} =& - N \sqrt g \left( R^{ij} - {\sfrac 1 2} g^{ij} R + \Lambda g^{ij} \right) 
+ \frac{N}{2\sqrt g} g^{ij} \left( p^{k\ell} p_{k\ell} - {\sfrac 1 2} p^2\right)- \frac{2 N}{\sqrt g} \left( p^{ik} p_k{}^j - {\sfrac 1 2} p \, p^{ij} \right) 
\\
&+ \nabla_k (p^{ij} \xi^k) - p^{ik} \nabla_k \xi^j - p^{kj} \nabla_k \xi^i 
+ \sqrt g \left( \nabla^i \nabla^j N - g^{ij} \Delta N \right)  \,.
\end{aligned}
\end{equation}
These equations depend on the metric and momenta $g_{ij}$, $p^{ij}$ which solve the constraint equations~(\ref{Constraints}), on a shift function $\xi_i$ which is fixed by the diffeomorphism gauge we choose (see below), and on a lapse function $N$ which we get by solving~(\ref{LFE})

\subsection{Lapse fixing equation}\label{LFEsec}

The lapse fixing equation~(\ref{LFE}) takes the following form after imposing the spherically-symmetric ansatz~(\ref{SphericalSymmetry}):
\begin{equation}
\begin{aligned}
\frac 1 {4\sigma \mu^2 } \left[6 f ^2 N  \mu^3 + 4 \sigma   \left(-\mu   \left(2 N'  \sigma' +N  \sigma'' \right)+N  \sigma'  \mu ' + N  \mu  ^3\right)  +N  \mu (\sigma')^2 \right.
\\
\left.+\sigma  ^2 \left(\mu   \left(3 N  s^2-8 N'' \right)+8 N'  \mu' \right) \right]   = 0 \,.
\end{aligned}
\end{equation}
Its solution is
\begin{equation}\label{SolutionLFE}
N = \frac{\sigma' }{2 \mu \sqrt{\sigma} } \left[ c_1 + c_2   \dashint_r^\infty \d z   \left( 4 \sigma - 8 m \sigma^{\frac 1 2} + A^2 \sigma^{-1}  \right)^{-3/2} \sigma'(z) \right] \,.
\end{equation}
where $\dashint$ is the Cauchy principal value integral (for reasons that will be clear below). The bottom integration boundary is chosen conventionally as $+\infty$. Changing its value just corresponds to shifting the integration constant $c_1$ by a constant value.  With the choice I made, $c_1$ is the lapse at infinity, because asymptotic flatness implies that $ \frac{\sigma' }{2 \mu \sqrt{\sigma} } \xrightarrow[r\to \infty]{} 1$.
 

In the singular phase  the two terms of $N$ in Eq.~(\ref{SolutionLFE}) are both regular and positive everywhere except at the origin. With the same coordinate choice as Sec.~\ref{SingularPhaseSec}, $\sigma = y^2$,  It is straightforward to see that, for small $y$, the term proportional to $c_1$ tends to $A/(2y^2)$, while the term proportional to $c_2$ tends to $\sim y^{-2}$ (because the integral is always finite, while the part multiplying it goes like $y^{-2}$), so both terms diverge. Therefore, calling $c_\st{crit}= \displaystyle -  \dashint_0^\infty \d z   \left( 4 \sigma - 8 m \sigma^{\frac 1 2} + A^2 \sigma^{-1}  \right)^{-3/2} \sigma'(z)$, the lapse goes to $+\infty$ the origin if $c_2 /c_1 > c_\st{crit} $, and it goes to $-\infty$ if $\displaystyle c_2 < c_\st{crit} $. At the boundary between these two regions, when $\displaystyle c_2 /c_1 = c_\st{crit} $, the lapse will vanish at the origin.


In the wormhole phase $\sigma'$ vanishes at the throat, and so does the polynomial $4 \sigma^2 - 8 m \sigma^{\frac 3 2} + A^2$, in such a way that its ratio with $(\sigma')^2$ remains finite. So one can see immediately that the term multiplying $c_1$ vanishes at the throat. Moreover, since the throat is an absolute minimum for $\sigma$,  $\sigma'$ changes sign upon crossing the throat, and so does the $c_1$ term in the lapse. On the other hand the $c_2$ term stays finite at the throat: in fact, inspecting Eq.~(\ref{DiffeogaugeSigma}) one can see that  $\sigma'$ vanishes at the throat \emph{linearly} in $r$ - meaning that $\sigma' \sim \text{\it const.} (r - m \chi_2/4)$ for $r \sim m \chi_2/4$, and similarly for  $(4 \sigma - 8 m \sigma^{\frac 1 2} + A^2 \sigma^{-1})^{1/2}$.
Then, in a neighborhood of the throat, the $c_2$ term goes like:
\begin{equation}
(r - m \chi_2/4)  \dashint_r^\infty  \d z \frac{\text{sign}(z - m \chi_2/4)}{(z - m \chi_2/4)^2}
= \text{\it const.}
\end{equation}
Notice that a key role is played by the principal value integral: when $r <m \chi_2/4$ the integration range crosses a nonintegrable pole of the integrand. But the principal value of the integral is well defined, because the left and right limit of the primitive for $r \to m \chi_2/4$ are equal in magnitude and opposite in sign.

The lapse goes to a constant for $r\to 0$, but with nonzero (positive) derivative.
This is characteristic of the lapse of an asymptotically flat region represented in inverted coordinates: $N \sim \alpha + \beta \, r  \xrightarrow[r \to r_0^2/r]{ }\alpha + \frac{ r_0^2 \beta}{r} $, which is the lapse of an asymptotically flat spacetime.

\begin{figure}[t!]%
\begin{center}
 \parbox{\wid}{\includegraphics[width=\wid]{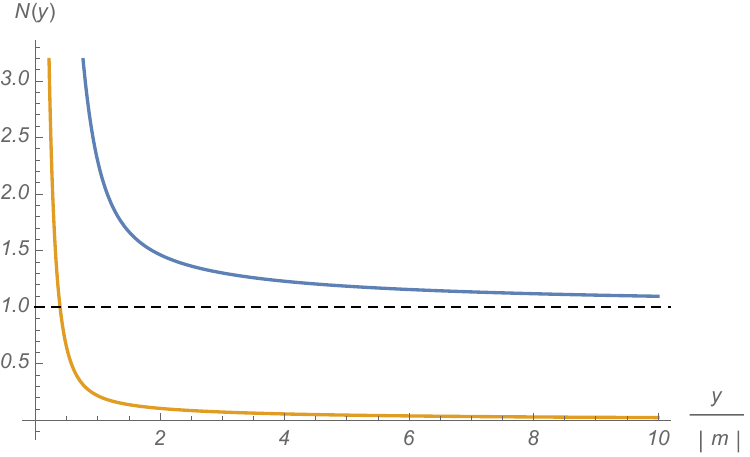}\figsubcap{a}}
 \hspace*{4pt}
 \parbox{\wid}{\includegraphics[width=\wid]{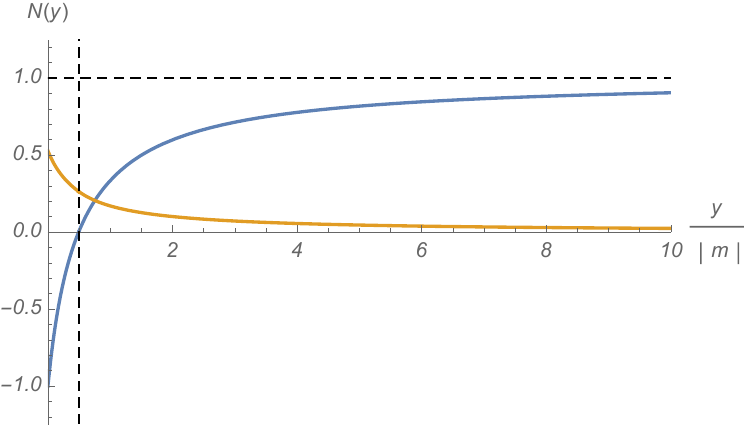}\figsubcap{b}}
 \caption{\small 
 Plot of the two terms of the lapse (the term proportional to $c_1$ in blue and the $c_2$ one in orange).  (a) For $C= \sqrt{4 \cdot 27} > \sqrt{27/16}$ and $m> 0$ (but the $m<0$ case is identical) (b)  For $C = 0.45<\sqrt{27/16}$.}
\end{center}
\end{figure}

In both phases we are in, the value of the parameter $c_2/c_1$ will determine the behaviour of the lapse at the origin. 
In the next section we will find the physical meaning of the parameter $c_2$, but for now we can draw some physical conclusions from the possible behaviours of the lapse.
In the cases in which the lapse is always positive, as in the singular phase when $c_2/c_1 > c_\st{crit}$ or in the wormhole case if $c_2$ is positive and sufficiently large, the non-backreacting matter propagating on this SD background will be able to reach every point of the spatial manifold, given sufficient maximal-slicing time. This means that, in the singular phase, it will be able to reach the singularity at the origin in a finite time (but the origin has to be `capped off' by matter in order to make sense of this case, so this is not a problem). In the wormhole phase it will be able to cross the throat and propagate freely towards the other asymptotic end. In the cases where the lapse goes through a zero, as in the singular case when $c_2/c_1 < c_\st{crit}$ or in the wormhole case when $c_2$ is negative or sufficiently small, there is an obstruction for matter to propagate towards the origin. The region where the lapse vanishes is an impenetrable barrier for matter, which cannot cross it in any maximal-slicing time. Of course this region can change its shape, depending on the time evolution of $c_2$, and can even disappear. As we will see below, everything depends on the time dependence of $A$.

%

\subsection{Equations of motion for the metric}\label{EqOfMotionSubSec}

Equation~(\ref{gdotEq}) look like this, under the spherically-symmetric ansatz:
\begin{equation} \label{HamiltonEqs_gdot}
\dot g_{ij}  =\left(\frac{2 f \mu^2 N}{\sigma} + 2 \mu \xi \mu' + 2 \mu^2 \xi'Ê\right) \delta^r{}_i \delta^r{}_j+  \left( \frac{\sigma s N}{\mu} + \xi \sigma' \right) \left( \delta^\theta{}_i\delta^\theta{}_j   + \delta^\phi{}_i \delta^\phi{}_j \sin^2 \theta \right) \,.
\end{equation}
This equation has two linearly independent components:
\begin{equation} \label{gdotrr}
2 \mu \dot \mu = \frac{2 f \mu^2 N}{\sigma} + 2 \mu \xi \mu' + 2 \mu^2 \xi'Ê \,,
\end{equation}
\begin{equation}\label{gdotthetatheta}
\dot \sigma  =  \frac{\sigma s N}{\mu} + \xi \sigma'  \,,
\end{equation}
Substituting the expressions for $s$ and $f$:
\begin{equation}
\begin{aligned}
&  2 \mu \dot \mu = \frac{2 A \mu^2 N}{\sigma^{\frac 3 2} } + 2 \mu \xi \mu' + 2 \mu^2 \xi'Ê \,,
&\qquad  &\dot \sigma  =   \xi \sigma'   - \frac{A}{\sigma^{\frac 1 2}} N \,,
\end{aligned}
\end{equation}
the right-hand side equations determines $\xi$:
\begin{equation}\label{SolutionForXi}
\begin{aligned}
&  \xi = \frac{\dot \sigma}{\sigma'} + \frac{A}{\sigma^{\frac 1 2} \sigma'} N(r) \,, 
\end{aligned}
\end{equation}
this equation fixes the shift once a diffeo gauge-fixing is chosen.  For example, if we choose the
gauge~(\ref{DiffeogaugeSigma}) we have $\dot \sigma = 4 \sigma^{3/4}m \, r^{-\frac 1 2} \dot\chi_2$, which fixes $\xi$. For our purposes we can keep $\xi$ unexpressed, because it will drop out of all equations.

Finally, replacing Eq.~(\ref{SolutionForXi}) into Eq~(\ref{gdotrr}) we get the following equation:
\begin{equation}\label{c2equalsAdot}
\begin{aligned}
-\left( 8 \sigma ^{3/2} \dot m + A (2 \dot A - c_2 Ê)  \right)\frac{\mu ^4 }{\sigma  \left(\sigma '\right)^2} = 0 \,,
\end{aligned}
\end{equation}
it is obvious that the only way to solve the above equation for each value of $r$ is to set
\begin{equation} 
\dot m = 0 \,, \qquad  c_2 =2 \dot A \,.
\end{equation}
We found out that $m$ is a conserved quantity. Moreover we fixed the remaining integration constant of $N$: $c_2$ is related to the time derivative of $A$.  Unfortunately, this means that we are not able to say much more about the behaviour of matter propagating on this solution than what was commented in Sec.~\ref{LFEsec}. The problem is that the system is underdetermined, and depends on data which cannot be determined from within the system (the time-dependence of $A$).

\subsection{The Misner--Sharp mass}

Eq.~(\ref{c2equalsAdot}) allowed us to discover a conserved quantity, $m$. This has a clear physical interpretation: it is the  value of the \emph{Misner--Sharp mass} of the system.
The general definition of  Misner--Sharp mass can be found for example in~\cite{GiuliniMcVittie}:
\begin{equation}
M_\st{MS} = \frac{\sqrt \sigma}{2} \left(  1 -  {^\st{(4)}g}^{\mu\nu}  \partial_\mu (\sqrt{\sigma}) \partial_\nu (\sqrt{\sigma}) \right) \,,
\end{equation}
where $\sqrt{\sigma} = \sqrt{g_{\theta\theta}}$ is the \emph{areal radius} coordinate of a spherically-symmetric metric. $M_\st{MS}$ is a 4-dimensional scalar which depends on the 4-metric. In our approach we can derive an effective 4D line element from the 3-metric $g_{ij}$, the momenta $p^{ij}$, the lapse $N$ calculated in~(\ref{LFE}) and the shift $\xi_i$ deduced from Eq.~(\ref{SolutionForXi}). The formula is:
\begin{equation}
{^\st{(4)}g}_{00} = - N^2 + g_{ij} \xi^i \xi_j \,, \qquad {^\st{(4)}g}_{0i} = g_{ij} \xi^j \,,  \qquad {^\st{(4)}g}_{ij} = g_{ij} \,.
\end{equation}
Of course this is just reverse-engineering the ADM construction, because we are exploiting the local equivalence between SD and GR.\footnote{In~\cite{Tim_Proceedings_TheoryCanada9} an interpretation of the above 4-metric in terms of the background spacetime that is `experienced' by weak matter fluctuation is given.}
Using the definition above, and the solutions we found for all the fields, we get the following expression for the Misner--Sharp mass:
\begin{equation}
M_\st{MS} 
=  \frac{\sqrt \sigma}{2} \left[ 1 -\frac{1}{4 \sigma }\left(\frac{(\sigma ')^2}{\mu ^2}-\frac{\left(\dot{\sigma }-\xi  \sigma '\right)^2}{N^2} \right) \right] \,,
\end{equation}
and, using the equations of motion~(\ref{gdotthetatheta}) for $\sigma$ and the maximal-slicing condition $s = - \mu f/\sigma$,
\begin{equation}
M_\st{MS} = \frac{\sqrt \sigma}{2}    -  \frac{(\sigma')^2}{8\sqrt{\sigma} \mu^2} + \frac{ f^2}{8\sqrt{\sigma}} = m \,,
\end{equation}
we get that $M_\st{MS}$ coincides with the expression~(\ref{DefinitionOfm}) defining the integration constant $m$. $M_\st{MS}$ is a concept of quasi-local mass, introduced in 1964~\cite{MisnerSharpMass}, which tries to capture this idea that in a spherically symmetric situation the only way mass-energy can escape from a sphere is by a physical flow of matter through the surface of the sphere. In vacuum regions it is a constant (as in our case, where $m$ is spatially constant), but if we had a spherically-symmetric distribution of matter it would be $r$-dependent. As $r\to \infty$, $M_\st{MS}$ reduces to the ADM mass of the system.

\subsection{Equations of motion for the momenta}

Replacing the spherically symmetric ansatz into Equation~(\ref{pdotEq}),
\begin{equation}\label{HamiltonEqs_pdot}
\begin{aligned}
\dot p^{ij}  =&    -\frac {\sin \theta} {4 \sigma \mu^3} \left[ 5 f^2 N \mu^2 + 
   4 \sigma (N' \sigma' - \mu^2 \xi f') + \
N (-4 \sigma  \mu^2 + (\sigma')^2) + 
   4 f \sigma  \mu (\xi  \mu' + \mu \xi')\right] \delta^i{}_r \delta^j{}_r \\
&+ \frac {\sin 
\theta} {4 \sigma^2 \mu^2} \left[ (f^2 N \mu^3 + N \mu (\sigma')^2 + 
   2 \sigma^2 (2 N' \mu' + \mu^2 (\xi s' + s \xi') - 
      2 \mu N'') \right.
      \\
  & ~~~~~~~~~~~~~   \left. - 2 \sigma (-N \sigma' \mu' + \mu (N' \sigma' + 
         N \sigma'')))\right] \left( \delta^i{}_\theta\delta^j{}_\theta   + \delta^i{}_\phi \delta^j{}_\phi \sin^{-2} \theta \right)  \,.
         \end{aligned}
\end{equation}
The two linearly independent  parts are
\begin{equation}
\dot{(f/\mu)} =    -\frac {1} {4 \sigma \mu^3} \left[ 5 f^2 N \mu^2 + 
   4 \sigma (N' \sigma' - \mu^2 \xi f') + \
N (-4 \sigma  \mu^2 + (\sigma')^2) + 
   4 f \sigma  \mu (\xi  \mu' + \mu \xi')\right] \,,
 \end{equation}
 \begin{equation}
 \begin{aligned}
 {\sfrac 1 2} \dot s = & \frac {1} {4 \sigma^2 \mu^2} \left[ (f^2 N \mu^3 + N \mu (\sigma')^2 + 
   2 \sigma^2 (2 N' \mu' + \mu^2 (\xi s' + s \xi') - 
      2 \mu N'') \right.
      \\
  & ~~~~~~~~~~~~~   \left. - 2 \sigma (-N \sigma' \mu' + \mu (N' \sigma' + 
         N \sigma'')))\right]  \,,
   \end{aligned}
\end{equation} 
substituting everything we can, we can put the above equations in the following form:
\begin{equation}
\begin{aligned}
\frac{\mu}{\sigma ^{3/2} \left(\sigma '\right)^2}  \left[\left(2 \dot A - c_2 \right) \left( 2 \sigma ^{3/2} \left(\sqrt{\sigma } - 2 m\right)+A^2 \right) - 2  \dot m  A \sigma ^{3/2} \right] = 0  \,,
\\
- \frac{\left(\sigma '\right)^4}{\sigma  \mu ^3}  \left[ 2 \dot m A +\left(2 m + \sqrt{\sigma }\right) \left(2 \dot A - c_2 \right)  \right] = 0 \,,
\end{aligned}
\end{equation}
these equations too are solved by $\dot m=0$ and  $c_2 = 2 \dot A$.

We solved all the available equations. Nothing fixed $A=A(t)$, so we are left with an underdetermination by one function of time. For what regards the system excluded its asymptotic boundary, any choice of $A(t)$ gives rise to a perfectly acceptable solution of maximal-sliced Shape Dynamics.

\section{Falloff conditions, boundary charges and Poincar\'e invariance}

In this section I will investigate first the physical meaning of the parameter $A$, which as we saw above determines the solution without being determined by anything internal to the system. Then I will discuss the invariance properties of the falloff conditions of the metric and the momenta: these are, in GR, used to select the correct boundary conditions for asymptotically flat spacetimes (\emph{i.e.} asymptotically Minkowski). In our case $A\neq 0$ breaks the Lorentz invariance of the boundary conditions: this implies that for a nonzero $A$ we have asymptotically flat spatial slices of Schwarzschild spacetime which introduce a preferred frame of reference at infinity. This further clarifies the physical role of $A$.

\subsection{The ADM charges}

The isometries of the boundary are associated to conserved quantities. In particular, to each asymptotic Killing vector (meaning vector fields $X^i$ that saisfy the Killing equation at the boundary, $\displaystyle \lim_{r \to \infty} (\nabla_i X_j + \nabla_j X_i )= 0$) there is an associated redundance of the diffeomorphism constraint. This is associated to a boundary term which is automatically conserved (because it is first-class wrt the total Hamiltonian), and is not set to zero by anything.
Our boundary admits the following Killing vectors:
\begin{equation}
\begin{aligned}
&\xi_x = \partial_x \,, & &
\xi_y = \partial_y \,, & &
 \xi_z = \partial_z \,,& \qquad
 & \text{(translations)}&
\\
&\chi_x = y \partial_z - z \partial_y \,,& &
\chi_y = z \partial_x - x \partial_z \,,& &
\chi_z = x \partial_y - y \partial_x \,,&
& \text{(rotations)} &
\end{aligned}
\end{equation}
so the associated charges can be interpreted as the linear and angular momentum of the boundary. In spherical coordinates the killing vectors read
\begin{equation}
\begin{aligned}
&\vec \xi_x = \left(\sin \theta \cos \phi, {\sfrac 1r} \cos \theta \cos \phi ,- {\sfrac 1 r} \csc \theta \sin \phi \right)\,, & &
\vec \chi_x = \left( 0,\sin \phi ,\cot \theta \cos \phi  \right) \,,&
\\
&\vec \xi_y = \left(\sin \theta \sin \phi,{\sfrac 1 r} \cos \theta \sin \phi,{\sfrac 1 r} \csc \theta \cos \phi \right) \,, & &
\vec \chi_y = \left( 0,-\cos \phi ,\cot \theta \sin \phi  \right) \,, &
\\
&\vec \xi_\st{z} = \left( \cos \theta,-{\sfrac 1 r} \sin \theta ,0\right)\,,&
&
\vec \chi_\st{z} = \left( 0,0,1 \right) \,.&
\end{aligned}
\end{equation}
 Then the linear momentum is~\cite{beig1987poincare}:
\begin{equation}
P_\st{A}  =  2  \int_{\partial \Sigma}  \,\xi_\st{A}^i \, p_{ij}  \, \d S^i =  \lim_{r\to \infty} \frac{A}{r} \int \sin \theta \left (\sin  \theta   \cos  \phi  , \sin  \theta   \sin  \phi  ,\cos  \theta   \right)_\st{A} \d \theta \d \phi = 0,
\end{equation}
where $\d S^i = \delta^i_r \d \theta \d \phi$.ÊThe angular momentum is
\begin{equation}
L_\st{A} = 2  \int_{\partial \Sigma}  \,\chi_\st{A}^i \, p_{ij}  \, \d S^i =2  \int_{\partial \Sigma}  \,\chi_\st{A}^r \, p_{rr}  \, \d \theta \d \phi = 0 \,.
\end{equation}

One can also calculate a sort of boundary charges associated to the \emph{conformal} Killing vectors, which satisfy the conformal Killing equation $\nabla_i X_j + \nabla_j X_i  - \frac 2 3 g_{ij} \nabla_k X^k = 0$. These charges are not conserved, because they are not associated to a first-class constraint. However, they can still be interpreted as momenta of the boundary.
The Killing vector fields of our metric, together with the regular Killing vectors, close an $SO(4,1)$ algebra (the 3-dimensional Euclidean conformal algebra). The conformal Killing vectors are the dilatation vector field $\varphi^i = x^i$,
and the three special conformal transformation vectors fields, $\kappa^i_\st{A} = 2 x^i x_\st{A} - \delta^i{}_\st{A}  x^i x_j$. Or, in spherical coordinates, $\vec \varphi = \left( r , 0 ,0\right)$, $\vec \kappa_x = \left( r^2 \cos \phi \sin \theta ,-r  \cos \theta  \cos \phi,r \csc \theta  \sin \phi  \right)$,
$\vec \kappa_y = \left( r^2 \sin \theta  \sin \phi ,-r  \cos \theta  \sin \phi ,-r \cos \phi \csc \theta \right)$ and $\vec \kappa_\st{z} = \left( r^2  \cos \theta ,r \sin \theta ,0\right)$.
The associated charges are:
\begin{equation}
D = 2  \int_{\partial \Sigma}  \,\varphi^i \, p_{ij}  \, \d S^i =4 \pi \, A  \,,
\qquad
K_\st{A} = 2  \int_{\partial \Sigma}  \,\kappa_\st{A}^i \, p_{ij}  \, \d S^i =  0\,.
\end{equation}
We see that the integration constant $A$ is proportional to the dilatation charge. We have finally found a physical interpretation of $A$: it is associated to the dilatational momentum of the matter at infinity. Indeed, if we couple a thin spherically symmetric shell of dust to Shape Dynamics, the solution inside the shell is identical to ours, and the integration constant $A$ is fixed by the diffeomorphism constraint to be proportional to the radial momentum of the shell.

\subsection{Asymptotic Poincar\'e (non-)invariance}

I want now to consider the falloff of the metric and momentum tensors at infinity,
$g_{ij} \sim \delta_{ij} + \frac 1 r \, \delta g_{ij}(\theta,\phi) + \mathcal O (r^{-2})$ and 
$p^{ij} \sim   \frac A r \, \delta p^{ij}(\theta,\phi) + \mathcal O (r^{-2})$, and study their
Poincar\'e invariance, following Beig and \'O Murchadha~\cite{beig1987poincare}.
To do this, one needs to introduce a linear combination of lapse scalars and shift vectors:
\begin{equation}
N = \alpha^0 + \left( \beta^x \,  \sin \theta \cos  \phi +\beta^z \, \sin \theta  \sin \phi  + \beta^y  \cos  \phi \right) \,,
\qquad 
\xi^i  =   \alpha^\st{A} \, \xi_\st{A}^i + \omega^\st{B} \, \chi_\st{A}^i \,,
\end{equation}
which generate every possible Poincar\'e transformation (at infinity): $\alpha^0$ is the parameter associated to time translations, $\beta^\st{A}$ generate a boost, $\alpha^\st{A}$ and $\omega^\st{A}$ generate, respectively, a translation and a rotation. Then one needs to calculate the Poisson bracket between the metric and the momenta with the total Hamiltonian:
\begin{equation}
H_\st{tot} = \int \d^3 x \left(N \, \mathcal H + \xi^i \, \mathcal H_i \right) \,,
\end{equation}
smeared with the lapse and the shift introduced here. If these Poisson brackets do not generate a higher-order term in $r$, then the falloff conditions are preserved by Poincar\'e transformations. For this calculation, it is important to notice that the falloff conditions written above in cartesian coordinates are different in spherical coordinates:
\begin{equation}
\begin{aligned}
& g_{rr}  = 1 + \mathcal O (1/r)  \,, & &  g_{r\theta} = g_{r\phi} /\sin^2\theta = r^2 + \mathcal O (1) \,, &
& g_{\theta\theta} =
 g_{\theta\phi}  =
 g_{\phi\phi} = \mathcal O (r)
\,, &\\
& p^{rr}  =  \mathcal O (1/r)  \,, & &  p^{r\theta} =    p^{r\phi} =   \mathcal O (1/r^2) \,, &
& p^{\theta\theta} =
 p^{\theta\phi}  =
 p^{\phi\phi} = \mathcal O (1/r^3)
\,. &
\end{aligned}
\end{equation}
The Poisson brackets with the momenta preserve the falloff conditions:
\begin{equation}
\begin{aligned}
& \{ p^{rr} ,   H_\st{tot} \} = \mathcal O (1/r)  \,, \qquad \{ p^{r\theta} ,   H_\st{tot} \}  =  \{ p^{r\phi} ,   H_\st{tot} \} =   \mathcal O (1/r^2)&
\\
&\{ p^{\theta\theta} ,   H_\st{tot} \} =  
 \{ p^{\theta\phi} ,   H_\st{tot} \} =  
\{ p^{\phi\phi} ,   H_\st{tot} \} =   \mathcal O (1/r^3) \,,&
\end{aligned}
\end{equation}
but the ones with the metric do not:
\begin{equation}\label{AbreaksLorentzEq}
\begin{aligned}
& \{ g_{rr} ,   H_\st{tot} \} =  2 A \, \left( \beta^x \, \sin \theta \cos \phi + \beta^y \, \sin \theta \sin \phi + \beta^z \cos \theta  \right) + \mathcal O (1/r)  \,,\
\\
& \{ g_{r\theta} ,   H_\st{tot} \} =  \mathcal O (1/r) \,, ~~~~
 \{ g_{r\phi} ,   H_\st{tot} \} =   \mathcal O (1/r)&
  \\
&\{ g_{\theta\theta} ,   H_\st{tot} \} =   \mathcal O (r)  \,, ~~~~
 \{ g_{\theta\phi} ,   H_\st{tot} \} =   \mathcal O (r)  \,, ~~~~
\{ g_{\phi\phi} ,   H_\st{tot} \} =   \mathcal O (r) \,,&
\end{aligned}
\end{equation}
in particular, it is the $A$ integration constant that makes the system not asymptotically invariant under boost. Any boost in any direction will break the falloff conditions. If the $A$ integration constant is set to zero, then, as is shown in~\cite{beig1987poincare}, the falloff conditions are invariant.

To prove that the falloff conditions are invariant under spacetime translations and space rotations it is not sufficient to show that they are preserved by these transformations. One has also to show that the parity of the leading-order components $ \delta g_{ij}(\theta,\phi)$ and $ \delta p^{ij}(\theta,\phi)$ are preserved, in the sense that also $ \{ g_{ij}(\theta,\phi)  H_\st{tot} \} $ and $ \{ p^{ij}(\theta,\phi)  H_\st{tot} \} $ have the same parity at leading order. In particular, under a parity transformation $\theta \to \pi -\theta$, $\phi \to \phi +\pi$, $g_{r\theta}$, $p^{r\theta}$, $g_{\phi\theta}$ and $p^{\phi\theta}$ are odd, and all the other components are even. An explicit calculation confirms that the parity of both the metric and the momenta are preserved at leading order.

This last calculation, in particular Eq.~(\ref{AbreaksLorentzEq}), shows that $A\neq0$ (and therefore, as shown above, the matter at infinity is expanding or contracting) introduces a preferred frame of reference, which breaks the Lorentz invariance of the boundary conditions.

\section{Discussion and conclusions}

The system is underdetermined. We do not have any more equations to appeal to, and nothing inside our system determines the dynamics of $A$: in principle it could be any function of time. This is a failure of Birkhoff's theorem: we only assumed spherical symmetry and no matter sources. This in GR is sufficient to uniquely fix the solution to be Schwarzschild's spacetime; however SD, upon imposing the same conditions, gives a family of solutions, one for each possible choice of $A=A(t)$. This underdetermination is cured by coupling gravity to spherically-symmetric sources with a compact support. Then the $A$ parameter of the metric outside of the support of matter turns out to be related, through the diffeomorphism constraint, to the dilatational momentum of the source. In~\cite{ThinShell} we show, using a thin spherical shell of dust as source, that $A$ is completely determined by the dynamics of the shell.

It may seem puzzling that the same assumptions give a unique solution in GR and a multiplicity of solutions in SD. However we must observe that every value of $A$ correspond to a particular maximal slice of Schwarzschild's spacetime. Maximal slices of  Schwarzschild have been studied, among others, by Estabrook \emph{et al.}~\cite{estabrook1973maximally} and Beig and \'O Murchadha~\cite{BeigNiall,beig2000maximal} (check Alcubierre's book~\cite{Alcubierre} for a review).
 \begin{center}
 \includegraphics[width=\textwidth]{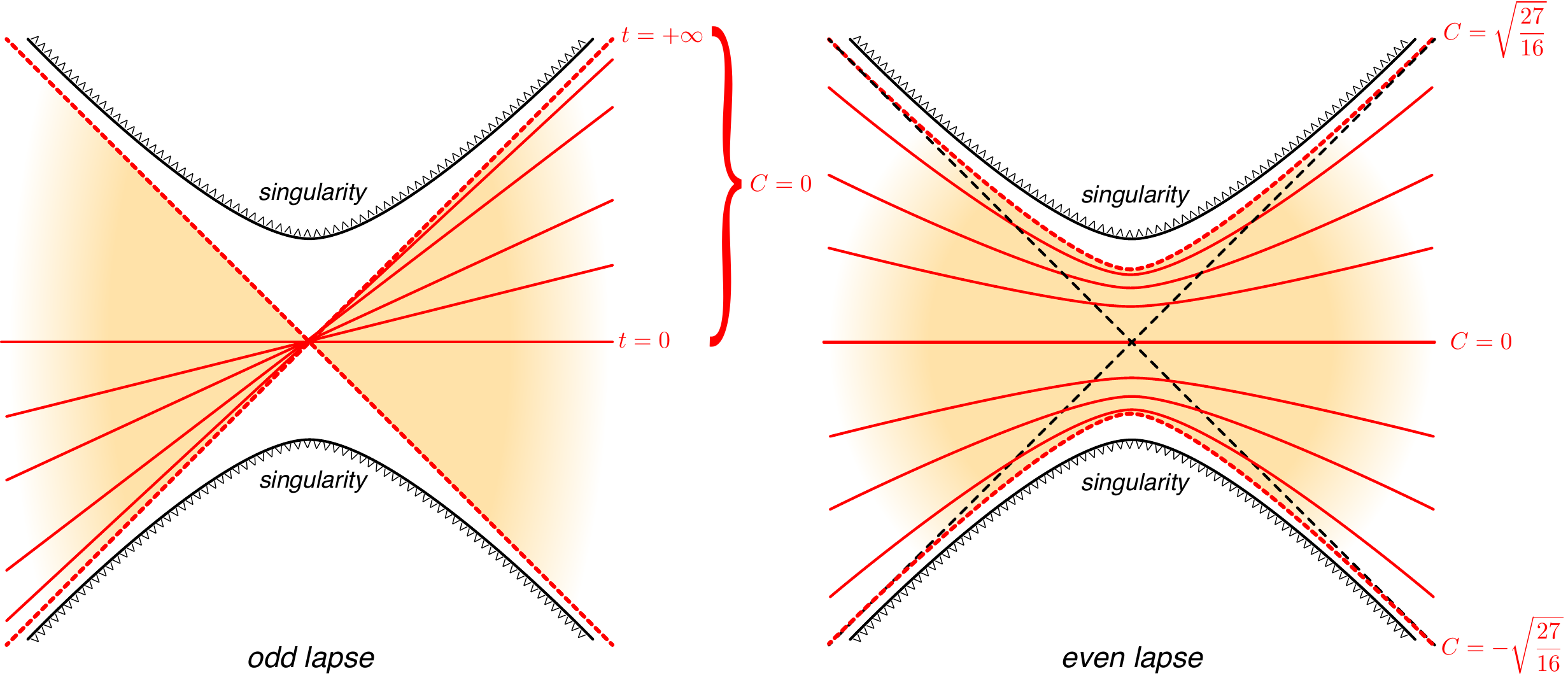}
 \end{center}
There are two kinds of maximal foliations of Schwarzschild spacetime  that include the  surface $T=0$ in Kruskal coordinates. One is the `odd-lapse' foliation, shown above on the left. The foliation covers only the right and left quadrants of Kruskal's diagram (the area in orange), never crossing the Schwarzschild horizon.
Each leave correspond to a constant-Schwarzschild-time ($t$) surface, each has two asymptotically flat boundaries, and crosses all the other leaves at the origin of Kruskal's diagram, where the lapse vanishes. On each of these slices the variable $C$ (and of course $A$) vanishes.
The even-lapse foliation covers the area between two constant-Schwarzshild-radius surfaces, $r_s =3m/2$ (in orange in the diagram), where $C^2$ reaches the critical value $27/16$. On each slice $C$ takes a different value, between the two critical values $C = \pm \sqrt{27/16}$. $C$ is a good time variable for this foliation (sometimes called the Estabrook--Wahlquist time), and the lapse vanishes nowhere. 
As I observed above, any choice of time-dependence for $A$ produces an acceptable solution of Shape Dynamics. Then the last remark on the maximal foliations of Schwarzschild implies that such solutions will be isometric to patches of Schwarzschild spacetime only in the intervals in which $A$ is monotonic.
Again, $A$ can only be determined by matter sources, as happens in a compact, dynamically closed universe.

\newpage

To conclude I will list the original technical results contained in this paper:
\begin{itemize}
\item
I showed that the most generic spherically-symmetric, asymptotically-flat solution of Shape Dynamics depends on two integration constants: $m$ and $A$.
\item
$m$ is conserved and coincides both with both the Misner--Sharp mass and the ADM mass.
\item
$A$ does not have to be conserved, and the quantity $C = A/(2m^2)$ determines whether there is a throat (a minimal-area sphere) or not. The throat is present only if $C<27/16$.
\item
I studied the metric in the case without a throat ($C > 27/16$), and it has a curvature singularity at the origin. However the conformal geometry is regular everywhere, because the Cotton--York density is zero. Moreover the inferred 4-metric is isometric to Schwarzschild.
\item
The case with a throat is qualitatively similar to what studied in~\cite{Birkhoff_SD}.
\item
I studied the falloff conditions for the metric and the momenta, and found that if $A \neq 0$ these conditions are not Lorentz-invariant. Moreover all the Poincar\'e ADM charges are zero, but the charge associated to dilatation is nonzero and proportional to $A$. The charge associated to special conformal transformations is zero.
\end{itemize}

\section*{Acknowledgements}

Perimeter Institute is supported by the Government of Canada through Industry Canada and by the Province of Ontario through the Ministry of Economic Development and Innovation. This research was also partly supported by grants from FQXi and the John Templeton Foundation.


\begin{thebibliography}{10}

\bibitem{FlaviosSDtutorial}
F.~Mercati, {\em Shape Dynamics: Relativity and Relationalism}.
\newblock Oxford Univ. Press, 2016.
\newblock \href{http://arxiv.org/abs/1409.0105}{{\ttfamily arXiv:1409.0105}}.
\newblock (preliminary version on the arXiv).

\bibitem{York1971}
J.~W. York, ``{Gravitational degrees of freedom and the initial-value
  problem},'' \href{http://dx.doi.org/10.1103/PhysRevLett.26.1656}{{\em Phys.
  Rev. Lett.} {\bfseries 26} (1971) 1656--1658}.

\bibitem{ChoquetBruhat73}
Y.~Choquet-Bruhat, ``Global solutions of the equations of constraints in
  general relativity on closed manifolds,'' {\em Symposia Math.} {\bfseries 12}
  (1973) 317.

\bibitem{Niall_73}
N.~O'Murchadha and J.~W. York, ``Existence and uniqueness of solutions of the
  Hamiltonian constraint of general relativity on compact manifolds,''
  \href{http://dx.doi.org/10.1063/1.1666225}{{\em J. Math. Phys.} {\bfseries
  14} (1973) 1551--1557}.

\bibitem{Birkhoff_SD}
H.~Gomes, ``A Birkhoff theorem for Shape Dynamics,'' {\em Class. Quantum Grav.}
  {\bfseries 31} (2014) 085008,
  \href{http://arxiv.org/abs/1305.0310}{{\ttfamily 1305.0310}}.
  \url{http://arxiv.org/abs/1305.0310}.

\bibitem{EinsteinChainOfCauses}
A.~Einstein, {\em The Meaning of Relativity}, p.~22.
\newblock Methuen, London, 1922.

\bibitem{beig1987poincare}
R.~Beig and N.~{\'O}~Murchadha, ``The Poincar{\'e} group as the symmetry group
  of canonical general relativity,'' {\em Annals of Physics} {\bfseries 174}
  no.~2, (1987) 463--498.

\bibitem{ThroughTheBigBang}
T.~A. Koslowski, F.~Mercati, and D.~Sloan, ``Relationalism Evolves the Universe
  Through the Big Bang,'' \href{http://arxiv.org/abs/1607.02460}{{\ttfamily
  1607.02460}}.

\bibitem{Gourg}
E.~Gourgoulhon, {\em 3+1 formalism in general relativity: bases of numerical
  relativity}.
\newblock Springer, 2012.

\bibitem{GiuliniMcVittie}
M.~Carrera and D.~Giulini, ``Generalization of McVittie's model for an
  inhomogeneity in a cosmological spacetime,''
  \href{http://dx.doi.org/10.1103/PhysRevD.81.043521}{{\em Phys. Rev.}
  {\bfseries D81} no.~4, (2010) },
  \href{http://arxiv.org/abs/0908.3101}{{\ttfamily gr-qc:0908.3101}}.

\bibitem{schoen1979proof}
R.~Schoen and S.-T. Yau, ``On the proof of the positive mass conjecture in
  general relativity,'' \href{http://dx.doi.org/10.1007/BF01940959}{{\em
  Communications in Mathematical Physics} {\bfseries 65} no.~1, (1979) 45--76}.

\bibitem{witten1981new}
E.~Witten, ``A new proof of the positive energy theorem,''
  \href{http://dx.doi.org/10.1007/BF01208277}{{\em Communications in
  Mathematical Physics} {\bfseries 80} no.~3, (1981) 381--402}.

\bibitem{Tim_Proceedings_TheoryCanada9}
T.~Koslowski, ``The shape dynamics description of gravity,''
  \href{http://dx.doi.org/10.1139/cjp-2015-0029}{{\em Can. J. Phys.} {\bfseries
  93} (2015) 956--962}, \href{http://arxiv.org/abs/1501.03007}{{\ttfamily
  arXiv:1501.03007 [gr-qc]}}.

\bibitem{MisnerSharpMass}
C.~W. Misner", ``Relativistic Equations for Adiabatic, Spherically Symmetric
  Gravitational Collapse,''
  \href{http://dx.doi.org/10.1103/PhysRev.136.B571}{{\em Physical Review}
  {\bfseries 136} no.~2B, (1964) B571--B576}.

\bibitem{ThinShell}
H.~Gomes, T.~Koslowski, F.~Mercati, and A.~Napoletano, ``Gravitational collapse
  of thin shells of dust in Shape Dynamics,''
  \href{http://arxiv.org/abs/1509.00833}{{\ttfamily arxiv:1509.00833}}.

\bibitem{estabrook1973maximally}
F.~Estabrook, H.~Wahlquist, S.~Christensen, B.~DeWitt, L.~Smarr, and E.~Tsiang,
  ``Maximally slicing a black hole,'' {\em Physical Review D} {\bfseries 7}
  no.~10, (1973) 2814.

\bibitem{BeigNiall}
R.~Beig and N.~{\'O}. Murchadha, ``Late time behavior of the maximal slicing of
  the Schwarzschild black hole,''
  \href{http://dx.doi.org/10.1103/PhysRevD.57.4728}{{\em Phys. Rev. D}
  {\bfseries 57} no.~8, (1998) 4728--4737},
  \href{http://arxiv.org/abs/gr-qc/9706046}{{\ttfamily arXiv:gr-qc/9706046}}.

\bibitem{beig2000maximal}
R.~Beig, ``The maximal slicing of a Schwarzschild black hole,'' {\em Ann.
  Phys.} {\bfseries 11} no.~5, (2000) 507--510,
  \href{http://arxiv.org/abs/gr-qc/0005078}{{\ttfamily arXiv:gr-qc/0005078}}.

\bibitem{Alcubierre}
M.~Alcubierre, {\em Introduction to 3+ 1 numerical relativity}.
\newblock Oxford Univ. Press, 2008.

\end{thebibliography}

\providecommand{\href}[2]{#2}\begingroup\raggedright\endgroup

\end{document}